\documentclass[11pt,a4paper]{article}

\usepackage[left=2.5cm, right=2.5cm, top=2cm, bottom=2cm]{geometry}

\usepackage{authblk}
\usepackage{graphicx} 
\usepackage{natbib}
\usepackage{amsmath,amssymb}
\usepackage{mathtools, amsthm}
\usepackage{tikz}
\usetikzlibrary{arrows.meta, positioning}
\usepackage{tikz-qtree}
\usepackage{enumitem}
\setlist[itemize]{label=\textbullet}
\usepackage{multirow}
\usepackage{comment}
\usepackage{array}
\usepackage[T1]{fontenc}
\usepackage{textcomp}
\usepackage{pgfplots}
\pgfplotsset{compat=1.18}
\usepgfplotslibrary{fillbetween}
\usepackage{iftex}
\usepackage{subcaption}
\usepackage[ruled,vlined,linesnumbered]{algorithm2e}
\usepackage[dvipsnames]{xcolor}
\definecolor{darkblue}{rgb}{0.0, 0.0, 0.55}
\definecolor{chicago-maroon}{RGB}{128,0,0}
\usepackage[bookmarks, colorlinks=true, plainpages = false, citecolor= darkblue, linkcolor = chicago-maroon, anchorcolor = red, urlcolor = darkblue]{hyperref}

\title{A Parameter-Centric View on Regression \\
\large{Discussion on ``Regression by Composition'' by Farewell, Daniel, Stensrud, and Huitfeldt}}
\author[1,2,3]{Jingxin Yan \thanks{jingxin.yan@mail.utoronto.ca}} 
\author[4]{Lin Liu \thanks{linliu@sjtu.edu.cn}}
\author[5]{Oliver Dukes \thanks{oliver.dukes@ugent.be}}
\author[2,3]{Qizhai Li \thanks{liqz@amss.ac.cn}}
\author[1]{Linbo Wang \thanks{linbo.wang@utoronto.ca}}
\affil[1]{Department of Statistical Sciences, University of Toronto, Toronto, Canada}
\affil[2]{State Key Laboratory of Mathematical Sciences, Academy of Mathematics and Systems Science, Chinese Academy of Sciences, Beijing, China}
\affil[3]{School of Mathematical Sciences, University of Chinese Academy of Sciences, Beijing, China}
\affil[4]{Institute of Natural Sciences, Shanghai Jiao Tong University, Shanghai, China}
\affil[5]{Department of  Mathematics, Computer Science, and Statistics, Ghent University, Ghent, Belgium}
\date{}

\usepackage{setspace}
\begin{document}
 
\maketitle
We congratulate the authors on a thought-provoking contribution. The paper addresses an important issue: although GLMs imply certain effect measures through appropriate link functions, many scientifically meaningful targets do not align naturally with standard GLM parameterizations. The authors therefore propose a more general modeling framework in which conditional distributions are constructed by composing group actions on distributions, rather than by relying on a single link function.

This raises a broader question: should we begin with a model or with the target parameter? We argue that if the primary goal is to model how a specific effect $\theta$ depends on covariates, then the natural starting point is $\theta$ itself. Within a parametric inferential framework, one could then construct a $\theta$-centric model with a parameterization that separates $\theta$ from suitable nuisance components. Ideally, this parameterization should be chosen to be variation independent of $\theta$, rather than embedding $\theta$ within a prespecified regression model whose coefficients happen to correspond to the desired effect.

To illustrate, suppose that the target effect is the conditional relative risk. A conventional approach uses a log link to model the outcome regression:
\begin{equation*}
\log P(Y = 1 \mid \text{trt}, \text{cov}) = \beta_0 + \beta_1 \text{trt} \cdot \text{cov} + \beta_2 \text{cov}.
\end{equation*}
Equivalently,
\begin{align*}
\log \frac{P(Y = 1 \mid \text{trt} = 1, \text{cov})}{P(Y = 1 \mid \text{trt} = 0, \text{cov})} &= \beta_1 \text{cov}, \\
\log P(Y = 1 \mid \text{trt} = 0, \text{cov}) &= \beta_0 + \beta_2 \text{cov}.
\end{align*}
The first equation models the target effect, whereas the second models a nuisance parameter, namely the baseline risk. In both standard GLM and regression-by-composition frameworks, this nuisance parameter is taken to be the baseline risk.

However, the baseline risk is variation dependent on many effect measures, including the relative risk and the risk difference. This dependence helps explain why Poisson regression is often viewed as unsuitable for modeling relative risks, why log-binomial models encounter constrained parameter spaces, and why linear models are criticized for modeling risk differences.

Once we move beyond prespecified regression models, it becomes possible to choose nuisance parameters that are variation independent of the target parameter. One example is the odds product \citep{richardson2017modeling}, which is variation independent of both the relative risk and the risk difference. When model fit is a concern, as illustrated in Figure 1 of \citet{farewell2026regression}, one can incorporate flexible nuisance modeling strategies, for example by modeling the odds product with a neural network and combining this with doubly robust procedures, or alternatively by adopting a fully nonparametric approach.

\bibliographystyle{apalike}
\bibliography{main.bib}

\end{document}